\documentclass[11pt]{article}
\pdfoutput=1

\usepackage{euscript}
\usepackage{amssymb}
\usepackage{amsfonts}
\usepackage{amsbsy}
\usepackage{epsfig}
\usepackage{amsthm}
\usepackage{amscd}
\usepackage{amstext}
\usepackage{verbatim}
\usepackage{amsmath}
\usepackage{cancel}
\usepackage{capt-of}
\usepackage{empheq}
\usepackage{hyperref}

\def\ben{\begin{equation}}
\def\een{\end{equation}}
\def\half{{\textstyle{\frac12}}}
\def\qtr{{\textstyle{\frac14}}}
\let\a=\alpha \let\b=\beta \let\g=\gamma \let\d=\delta 
   \let\k=\kappa
\let\l=\lambda     \let\r=\rho
\let\s=\sigma \let\t=\tau

\let\pa=\partial

\newcommand{\hph}[1]{\hphantom{#1}}

\newcommand{\ba}{\begin{align}}
\newcommand{\ea}{\end{align}}

\newcommand{\bi}{\begin{itemize}}
\newcommand{\ei}{\end{itemize}}

\newcommand{\dd}{{\rm d}}
\newcommand{\DD}{{\rm D}}

\newcommand{\Eqref}[1]{Eq. (\ref{eq:#1})}
\renewcommand{\eqref}[1]{eq. (\ref{eq:#1})}

\def\be{\begin{equation}}
\def\ee{\end{equation}}
\def\beq{\begin{equation}}
\def\eeq{\end{equation}}

\def\dalemb#1#2{{\vbox{\hrule height .#2pt
       \hbox{\vrule width.#2pt height#1pt \kern#1pt
               \vrule width.#2pt}
       \hrule height.#2pt}}}

\newcommand{\bea}{\begin{eqnarray}}
\newcommand{\eea}{\end{eqnarray}}

\def\R{{{\Bbb R}}}

\def\Z{{{\Bbb Z}}}

\newcommand{\rchi}{{\mathpalette\irchi\relax}}
\newcommand{\irchi}[2]{\raisebox{\depth}{$#1\chi$}}

\begin{document}

\begin{center}

{ \Large {\bf
Clumping and quantum order: \\
Quantum gravitational dynamics of NUT charge
}}

\vspace{1cm}

Sean A.~Hartnoll and David M.~Ramirez

\vspace{1cm}

{\small
{\it Department of Physics, Stanford University, \\
Stanford, CA 94305-4060, USA }}

\vspace{.25cm} {\small \texttt{hartnoll@stanford.edu},
  \texttt{dram@stanford.edu}}

\vspace{1.6cm}

\end{center}

\begin{abstract}

Gravitational instantons with NUT charge are magnetic monopoles upon dimensional reduction. We determine whether NUT charge can proliferate via the Polyakov mechanism and partially screen gravitational interactions. In semiclassical Einstein gravity, Taub-NUT instantons experience a universal attractive force in the path integral that prevents proliferation. This attraction further leads to semiclassical clumping instabilities, similar to the known instabilities of hot flat space and the Kaluza-Klein vacuum. Beyond pure Einstein gravity, NUT proliferation depends on the following question: is the mass of a gravitational instanton in the theory always greater than its NUT charge? Using spinorial methods we show that the answer to this question is `yes' if all matter fields obey a natural Euclidean energy condition. Therefore, the attractive force between instantons in the path integral wins out and gravity is dynamically protected against screening. Semiclassical gravity with a compactified circle can be self-consistently quantum ordered, at the cost of suffering from clumping instabilities.

\end{abstract}

\pagebreak
\setcounter{page}{1}

\section{Ideological preamble}

In this paper we consider aspects of the emergence of gravity, and the existence of massless gravitons in particular, from the viewpoint of the set of ideas referred to as `quantum order' in condensed matter physics \cite{Wen:2002zz,Wen:2004ym}. The robustness of massless excitations against becoming gapped needs to be explained, not assumed. Previous works have attempted microscopic derivations of the emergence of gravitons from constrained lattice models, by analogy to cases in which the emergence of photons in condensed matter systems has been understood. See for instance \cite{Lee:2006gp, Gu:2009jh, Xu:2010eg, Lee:2011wx}. Instead, we will approach the question from a semiclassical low energy effective field theory perspective. The effective field theory approach can constrain microscopic models. For instance, in our proof below of a positive mass theorem for gravitational instantons, the matter will need to satisfy a local energy condition.

In this paper we will also touch upon the nature of `spacetime foam' in quantum gravity. Most previous semiclassical studies of spacetime foam have focussed on counting saddle point solutions to the Einstein equations, e.g. \cite{Hawking:1979zw, Carlip:1997aa}. While multi-instanton configurations that are only approximate saddle points may appear to be suppressed, in fact the entropic proliferation of such configurations can lead to important qualitative effects. The canonical non-gravitational example that illustrates this fact is the confinement of $U(1)$ gauge fields in 2+1 dimensions due to the proliferation of magnetic monopoles \cite{Polyakov:1975rs}. Similar entropically-driven dynamics also arises in the unbinding of vortices across the Kosterlitz-Thouless transition \cite{KT}. The possibility of the proliferation of gravitational instantons with NUT charge was raised by Gross in \cite{Gross:1983mq}, but does not appear to have been further discussed.\footnote{Instanton proliferation in quantum gravity has been discussed in the context of Euclidean wormhole dynamics, see e.g. \cite{Gupta:1988nb, Gupta:1989bs}. In these cases the dominant physics involved in the instanton interactions and proliferation is not gravitational.} Some of the discussion in that paper anticipates our concerns in this work.

Instantons are localized excitations in the Euclidean path integral that carry a topological charge. They are field configurations valued in a nontrivial vector bundle over the spacetime. While exact multi-instanton solutions can exist, often the superposition of well separated instantons is not a solution to the Euclidean equations of motion. This is familiar from e.g. tunneling in quantum mechanics. This fact can be expressed as the presence of `forces' between instantons. Multi-instanton configurations are therefore penalized via an `energy cost' in the path integral. However, if the inter-instanton forces fall off sufficiently quickly at large distances then the entropic gain in accessing a multi-instanton configuration space can overwhelm the energy cost. In this case we say that the instantons `proliferate' and field configurations valued in nontrivial vector bundles may make a significant contribution to the path integral. In contrast, if for whatever reason instantons configurations are suppressed -- and furthermore there is no other dynamics, not necessarily described by an instanton gas, that leads to topological charge in the path integral \cite{Witten:1978bc} -- then the path integral exhibits a certain topological simplicity. We can express this fact by calling the system `quantum ordered'. This is our effective field theory perspective on quantum order. It is not totally satisfactory because the topological complexity of the path integral may not be invariant under dualities that amount to nontrivial field redefinitions.

Quantum order is necessary to protect the masslessness of 2+1 dimensional photons. As we review in section \ref{sec:polyakov}, Polyakov has famously shown that in a three dimensional $U(1)$ gauge theory, Debye screening of instanton interactions in a gas of instanton and anti-instantons can lead to instanton proliferation and give a mass to the photon \cite{Polyakov:1975rs}. One way in which quantum order can manifest itself and be understood from effective field theory in these cases is the fact that monopole proliferation can be suppressed by the presence of massless fermions, e.g. \cite{hermele,Unsal:2008sc, Dyer:2013fja}.

An important class of gravitational instantons carry NUT charge \cite{Gibbons:1979xm}. As we also recall below, these instantons are intimately connected to magnetic monopoles upon dimensional reduction on a circle \cite{Sorkin:1983ns, Gross:1983hb,Gross:1983mq}. Two circumstances where such dimensional reduction can occur is at finite temperature, in which case one can reduce on the Euclidean thermal circle, and in a Kaluza-Klein universe in which a spatial dimension is compactified. In these cases, NUT charged gravitational instantons have the potential to proliferate via the Polyakov mechanism and partially screen the long range gravitational interaction \cite{Gross:1983mq}. The most important part of this paper is the derivation of a new positive mass theorem for gravitational instantons in section \ref{sec:theorem}. We show that if the matter fields obey a natural Euclidean local energy condition, then the mass of a gravitational instanton is greater than its NUT charge. This statement will be seen to imply that
proliferation of NUT charge does not occur. In this case, the effective low energy quantum gravity theory is self-consistently quantum ordered (at least against the proliferation of NUT charge). In warming up to the general discussion we first consider the case of pure Einstein gravity in section \ref{sec:grav}. We also consider Einstein-Maxwell theory in detail in section \ref{sec:EM}, as dressing the NUT centers with electromagnetic fields introduces extra inter-instanton forces into the computation.

It is well known that both finite temperature gravity and Kaluza-Klein compactifications on a circle suffer from perturbative \cite{Appelquist:1982zs, Rubin:1983ap, Gross:1982cv} and non-perturbative instabilities \cite{Gross:1982cv, Witten:1981gj}. In the finite temperature case, that we shall mostly have in mind in this paper, these are clumping instabilities due to the universal attraction of the gravitational force. One important message of this paper is that it is precisely this same clumping physics that prevents the gravitational instantons from proliferating and thereby protects the massless graviton. The question of proliferation may appear moot given that the setups at hand are ultimately unstable. However, if new nonperturbative dynamics (proliferation) occurs, we may expect to find regimes in which this dynamics can compete with or overcome the various instabilities of the system. Relatedly, as we note in the discussion section, in circumstances in which gravitational clumping can be tamed, for instance in an expanding universe, one might expect the possibility of gravitational instanton proliferation to arise.

\section{Polyakov's computation}
\label{sec:polyakov}

We quickly review Polyakov's computation \cite{Polyakov:1975rs, Polyakov:1976fu} in order to establish notation and concepts that we will use below.
Monopole solutions of $U(1)$ gauge fields in 3 spacetime dimensions describe instanonic processes that create or annihilate vorticity. For the purposes of long distance physics the theory can be taken to have Euclidean action
\be\label{eq:qed}
S_E = \frac{1}{4 e_3^2} \int d^3x f^2 \,,
\ee
where $f = da$ is the Maxwell field strength. The Dirac monopole solution has
\be\label{eq:dirac}
a = \frac{\sigma}{2} \cos \theta \, d\phi \,.
\ee
here $\sigma = \pm 1$ while charge quantization fixed the coefficient. The action of a single monopole has a short distance divergence and the theory (\ref{eq:qed}) must be supplemented with a UV completion such as a lattice or an embedding into a non-Abelian theory. This leads to a single instanton action $s_o$ that is non-universal.

The interaction between two well separated instantons is however universal. Because multi-instanton configurations are not exact solutions to the equations of motion, we have to define what we mean by a multi-instanton configuration. Therefore we do the following. If the separation between the instantons is parametrically larger than the core size, then, for the purposes of determining the interaction, the core may be replaced by a delta function source for the magnetic field. A multi-instanton configuration will then be characterized by a fixed number of delta functions sources and we do the path integral over the magnetic field subject to the presence of sources. As is very well-known, this is equivalent to inducing a $1/r$ interaction between the sources.

To introduce the magnetic sources it is convenient to S-dualize. Up to overall normalization of the partition function, we have
\bea
\lefteqn{\int {\mathcal D}a \, e^{-  \int d^3x f^2/4 e_3^2} = \int {\mathcal D}f \, \delta(df) \, e^{-\int d^3x f^2/4 e_3^2}} \nonumber \\
&& =  \int {\mathcal D}f \, {\mathcal D} \Phi \, e^{ - \int d^3x f^2/4 e_3^2+ i/(2 e_3) \int d^3x \, \epsilon^{a b c} \Phi \pa_{a} f_{bc}} 
 =  \int {\mathcal D} \Phi \, e^{ - \frac{1}{2} \int d^3x \, (\pa \Phi)^2} \,. \label{eq:manip}
\eea
The original field strength is related to the dual scalar $\Phi$ through
\be\label{eq:dual}
f^{ab} = - i e_3 \epsilon^{cab} \, \pa_c \Phi \,.
\ee
The Dirac monopole solution (\ref{eq:dirac}) is seen to satisfy
\be
\nabla^2 \Phi = - \sigma \frac{2 \pi i}{e_3} \delta^{(3)}(x) \,.
\ee

The partition function in the presence of monopole and antimonopole sources can therefore be written as
\bea
Z & = & \sum_{m,n=0}^\infty \int \left[\prod_{i=1}^m d^3x_i \prod_{j=1}^n d^3y_j\right] \frac{e^{-(m+n) s_o}}{m! \, n!} \int {\mathcal D} \Phi \, e^{ - \frac{1}{2} \int d^3x \, (\pa \Phi)^2} e^{i 2 \pi [\sum_i \Phi(x_i) - \sum_j \Phi(y_j) ]/e_3} \nonumber \\
& = & \int {\mathcal D} \Phi \, e^{ - \widetilde S_E[\Phi]} \,. \label{eq:sumsources}
\eea
where the scalar field action
\be
\widetilde S_E[\Phi] = \int d^3x \left[\frac{1}{2} (\pa \Phi)^2 - 2 \, e^{-s_o} \cos (2 \pi \Phi/e_3) \right] \,.
\label{eq:polyakov}
\ee
We can see that this theory now describes a gapped photon by adding a current source for the Maxwell field. It is easiest to add a source $j^{ab}$ directly for the field strength $f_{ab}$. Following the manipulations in (\ref{eq:manip}) we have
\be
Z[j] = \int {\mathcal D}a \, e^{-  \int d^3x f^2/4 e_3^2 + \int d^3x f_{ab} j^{ab} } = \int {\mathcal D} \Phi \, e^{ - \frac{1}{2} \int d^3x \, (\pa \Phi + i e_3 \, \star j )^2} \,,
\ee
where $(\star j)_a = \epsilon_{a b c} j^{b c}$. Therefore the two point function becomes
\be
\Big\langle f_{pq}(x) f_{rs}(y) \Big\rangle = \left. \frac{\delta^2 Z[j]}{\delta j^{pq}(x) \delta j^{rs}(y)} \right|_{j = 0} =  e_3^2 \, \epsilon^a{}_{p q} \epsilon^b{}_{r s} \Big(\delta_{a b} \delta(x-y) - \Big\langle \pa_a \Phi(x) \pa_b \Phi(y) \Big\rangle \Big) \,. \label{eq:position}
\ee
Now we note that because the cosine interaction in (\ref{eq:polyakov}) is relevant, then at low energies the field will be sharply localized in a minimum of the potential. Therefore we can expand the cosine in powers of $\Phi$ and to lowest order we obtain a free theory for $\Phi$ with a mass $M^2 = 2 (2 \pi/e_3)^2 e^{- s_o}$. Taking the Fourier transform of (\ref{eq:position}) and Wick contracting we obtain
\be
\Big\langle f_{pq} f_{rs} \Big\rangle(p) = e^2_3 \, \epsilon^a{}_{p q} \epsilon^b{}_{r s} \Big(\delta_{a b} - \frac{p_a p_b}{p^2 + M^2} \Big) \,,
\ee
wherein we see that the photon propagator is gapped \cite{Polyakov:1975rs, Polyakov:1976fu}. The proliferation of a magnetically charged instanton gas in the path integral has Debye screened the (dualized) Maxwell field.

The upshot of this discussion is that the proliferation of topologically nontrivial field configurations in the path integral has gapped the would-be massless gauge field. The preservation of a massless gauge field -- the existence of a `critical phase' in condensed matter parlance -- requires a certain `quantum order' to be established. In the present context this would mean that topologically nontrivial configurations are suppressed in the path integral. For instance, it is known that monopole proliferation can be suppressed by the presence of massless fermions. For a discussions of fermions and monopole proliferation very much in the spirit of the present paper see \cite{hermele, Unsal:2008sc}.

\section{Einstein gravity}
\label{sec:grav}

In the remainder of the paper we will determine whether Polyakov-like proliferation can occur in gravitational theories. Consider first the case of pure Einstein gravity. The Euclidean action is
\be\label{eq:se}
S_E = -\frac{1}{2 \k^2} \int d^4x \sqrt{g} R - \frac{1}{\k^2} \int d^3x \sqrt{\gamma} K \,.
\ee
Here the trace of the extrinsic curvature is $K = \gamma^{\mu\nu} \nabla_\mu n_\nu$, with $n$ the outward pointing unit normal at infinity and $\g$ is the induced metric on the boundary. The single-centered instantons of interest are given by the Euclidean Taub-NUT metrics \cite{gi}. These are described by the following solution to the vacuum Einstein equations $R_{\mu\nu}=0$:
\be\label{eq:tn2}
ds^2 = U^{-1} \left(d \tau + \sigma \, n  \cos \theta d\phi \right)^2 + U \left(dr^2 + r^2 \left[d\theta^2 + \sin^2 \theta \, d\phi^2 \right] \right) \,,
\ee
with
\be\label{eq:period}
\tau \sim \t + \frac{1}{T} \,, \qquad n = \frac{1}{4 \pi T} \,, \qquad \sigma = \pm 1\,, \qquad U = 1 + \frac{n}{r} \,.
\ee
We will mostly think of $\tau$ as the Euclidean time direction, and therefore the identification $\tau \sim \t + 1/T$ in (\ref{eq:period}) indicates that these instantons contribute to the quantum gravity partition function at a finite temperature $T$. However, one can also think of $\tau$ as a spatial direction that has been compactified on a circle at zero temperature. This changes the interpretation of various instabilities that exist, as we will recall below.

The on-shell action of the single-centered instanton is given by
\be\label{eq:tnaction}
s_o = \frac{1}{2 \, \k^2 T^2} \,.
\ee
Unlike in the Dirac monopole case, this finite action is obtained within the universal Einstein theory (\ref{eq:se}). There are no source terms in the equations of motion. Processes mediated by these instantons therefore exist independently of the UV completion of the theory. The evaluation of the action of the Taub-NUT instanton has a slightly involved history. The validity of the initial computation \cite{Gibbons:1979nf} appeared questionable \cite{Hunter:1998qe} because, in order to obtain a finite result, the action of empty hot flat space was subtracted from the Taub-NUT action. The boundary metrics inherited by a cutoff surface at large radius were topologically distinct in the two cases. However, the result of \cite{Gibbons:1979nf} was recovered in \cite{Emparan:1999pm} by taking the flat space limit of the Taub-NUT solution in AdS spacetime. In asymptotically AdS spacetimes, holographic renormalization allows thermodynamic quantities to be computed without subtracting arbitrary reference spacetimes. This is the result we have quoted in (\ref{eq:tnaction}). A further argument for the validity of this result is to consider a well-separated NUT-anti-NUT spacetime. We will consider such configurations in detail shorty in order to obtain the multi-instanton contribution to the path integral. While not an exact solution to the equations of motion, it is an asymptotically flat spacetime and the action can be computed via subtracting the hot flat space answer. The result obtained to leading order at large separations is twice (\ref{eq:tnaction}).

In order to describe the interactions between well separated instantons, it is helpful to perform a dimensional reduction. This will also enable us to make a direct connection to the physics of monopoles outlined in the previous section \ref{sec:polyakov}. It is well known that dimensional reduction of NUT charge leads to magnetic monopoles \cite{Sorkin:1983ns, Gross:1983hb}. Here we are mainly thinking of reducing along the Euclidean time circle itself. The idea that NUT charge might proliferate via the Polyakov mechanism was first explored in \cite{Gross:1983mq}. The multi-instanton configurations we consider will fit into the ansatz
\be\label{eq:metricreduce}
ds^2 = e^{-\phi} \left(d\tau + a \right)^2 + e^\phi dx^2_3 \,.
\ee
Here $dx^2_3$ is the flat metric in three dimensions and $\phi, a$ do not depend on the Euclidean time coordinate $\tau$.
The action (\ref{eq:se}) evaluated on such configurations takes the form
\bea
S_E & = & \frac{1}{2 \k^2 \, T} \int d^3x \left(\frac{1}{4} e^{-2\phi} f^2 + \frac{1}{2} \left( \nabla \phi\right)^2 + \nabla^2 \phi \right) - \frac{1}{2 \k^2 \, T} \int dS \cdot \nabla \phi  \\
& = &  \frac{1}{2 \k^2 \, T} \int d^3x \left(\frac{1}{4} e^{-2\phi} f^2 + \frac{1}{2} \left( \nabla \phi\right)^2 \right) \,. \label{eq:phif}
\eea
Here $f = da$ is a three dimensional field strength. The final term is the Gibbons-Hawking term, in which we dropped a constant term. The boundary is a sphere at large radius.
The leading interaction between well separated instantons is dominated by the regime away from the centers in which the potentials are small. We can therefore linearize the above action in $\phi$ and $f$. Linearizing and then S-dualizing the field strength as we did in section \ref{sec:polyakov}, the action with a single source at the origin is given by
\be
\int d^3x \left[\frac{1}{2 \a_o^2} \left( \left(\pa \Phi \right)^2 + \left(\pa \Psi \right)^2 \right) - \delta^{(3)}(x) \left(i \sigma \Phi(x) + \Psi(x) \right) \right] \,. \label{eq:phipsi}
\ee
Here $\Phi$ is the dual to $f$, and we rescaled $\phi$ to $\Psi$. We have chosen the normalization of $\Phi$ and $\Psi$ so that the source terms in (\ref{eq:phipsi}) come with no prefactors. This normalization leads to the coefficient of the kinetic terms in (\ref{eq:phipsi}) being given in terms of
\be
\a_o^2 = \frac{1}{2 \k^2 T^3} \,.
\ee
Summing over sources as in (\ref{eq:sumsources}) then gives the effective action
\be
\widetilde S_E[\Phi,\Psi] = \int d^3x \left[\frac{1}{2 \a_o^2} \left( \left(\pa \Phi \right)^2 + \left(\pa \Psi \right)^2 \right)  - 2 \, e^{-s_o} e^{\Psi} \cos \Phi \right] \,.
\label{eq:nuteff}
\ee

Comparing the above effective action with the three dimensional electrodynamics case (\ref{eq:polyakov}) of the previous section, we see that the key difference is that the extra field $\Psi$ has resulted in an action that is now unbounded below as $\Psi \to \infty$. This behavior is not due to the inherent unboundedness of the conformal mode in the Euclidean Einstein action. Rather, the physics is straightforward to intuit: in the electrodynamics case the force between instantons could be positive or negative depending on the relative magnetic charge of the centers. This allows the inter-instanton force to be dynamically screened. With the forces screened, instanton proliferation is possible due to the entropic dominance of well separated instantons. In the gravitational case, the magnetic force is supplemented by a purely attractive gravitational force between all instantons. This gravitational attraction precisely cancels out the repulsive magnetic forces and adds to the attractive magnetic forces. There is therefore a net attraction overall and the gas of instantons anti-screens rather than screens itself \cite{Gross:1983mq}. Instead of proliferation, this leads to a clumping instability in which the action can be made arbitrarily negative as all the instantons clump together. The dilute gas approximation will of course break down in this process. The main conclusion is that proliferation does not occur.

The discussion in the previous paragraph is made more precise as follows. To compute the density $\rho(x) \equiv \sum_i \delta(x-x_i)$ of instantons, we can introduce a source into the partition function: $\exp \left[{\int d^3x \rho(x) J(x)} \right]$. Once the sum over instantons is performed, as in (\ref{eq:sumsources}), it is easily seen that adding the source term amounts to a shift $\Psi \to \Psi + J$ in the potential term of the effective action (\ref{eq:nuteff}). It follows that the expectation value of the density is
\be
\langle \rho \rangle = \left. \frac{\delta Z}{\delta J} \right|_{J = 0} = \left\langle 2 e^{-s_o} e^{\Psi} \cos \Phi \right\rangle \,.
\ee
We therefore see that as the effective potential becomes arbitrarily negative, the instanton density diverges. This is the clumping instability.

It is well known that gravitating systems at finite temperature are unstable towards clumping \cite{Gross:1982cv}. The clumping instability we have uncovered is a cousin of the well-known instabilities. In the semiclassical regime where $e^{-s_o}$ is very small, we can compute the decay rate of hot flat space to pair production of NUT charges. The decay rate is given by the imaginary part of the free energy of the system. The semiclassical instability discovered in \cite{Gross:1982cv} was mediated by the Schwarzschild instanton and the imaginary part of the free energy originated in a negative Euclidean mode about the Schwarzschild instanton. The single-centered Taub-NUT metric does not have a negative mode, as is guaranteed by the self-duality of the Riemann curvature tensor \cite{Hawking:1979zs,Young:1983dn}. The negative mode in our case is instead a fluctuation of the two-centered geometry. The contribution of a NUT-anti-NUT pair to the partition function is
\begin{align}
Z_{+-} & = \int d^3x d^3y \, e^{-2 s_o} \int {\mathcal D} \Phi \int {\mathcal D} \Psi e^{- \frac{1}{2\a_o^2} \int d^3x  \left( \left(\pa \Phi \right)^2 + \left(\pa \Psi \right)^2 \right)} e^{i \Phi(x) + \Psi(x) } e^{- i \Phi(y) + \Psi(y)} \\
& = e^{-2 s_o} \, Z_o \int d^3x d^3y \, e^{\a_o^2/(2 \pi |x - y|)} \\
& = V \, e^{-2 s_o} \, Z_o \, 4 \pi \int_0^\infty dr r^2 e^{\a_o^2/(2 \pi r)} \,. \label{eq:integral}
\end{align}
Here $V$ is the spatial volume and $Z_o$ is the partition function with no sources. The final integral in (\ref{eq:integral}) is clearly divergent. This is the manifestation at the two-instanton level of the fact that the full effective action (\ref{eq:nuteff}) is unbounded below due to the net attraction between the instantons. In the semiclassical limit however, we can extract a decay rate from (\ref{eq:integral}) for the production of NUT-anti-NUT pairs in hot flat space. One way to extract a well-defined imaginary part from the integral is, following \cite{Bogomolny:1980ur}, to analytically continue $\a_o^2 \to - \a_o^2$, do the integral (with a large cutoff on $r$ that is taken to infinity), and then continue back. More physically, we can estimate the decay rate by placing the $r^2$ in (\ref{eq:integral}) into the exponent and finding the point of unstable equilibrium between the attractive inter-instanton force and the entropic logarithmic repulsion between the instantons: $r_\star = \a_o^2/4 \pi$. The negative mode about this unstable saddle point gives the decay rate per unit volume. Writing $r = r_\star + \delta r$ and perturbing about the saddle point
\be\label{eq:gamma}
\Gamma \sim e^{-2 s_o} \text{Im} \int_{-\infty}^\infty d \d r e^{16 \pi^2 (\delta r)^2/\a_o^2} \sim e^{-2 s_o} \,.
\ee
We have dropped non-exponential terms as these are sensitive to the precise measure on the single-instanton moduli space, that we have not computed. The single-center Taub-NUT action (\ref{eq:tnaction}) is in fact also the action of the Schwarzschild instanton. Therefore, the rate of instability found in (\ref{eq:gamma}) is subleading compared to the semiclassical instability of hot flat space to the nucleation of black holes, which occurs at the rate $e^{-s_o}$ \cite{Gross:1982cv}.

The Euclidean Taub-NUT solution (\ref{eq:tn2}) is a special case of a more general family of solutions to pure Einstein gravity in which a `magnetic' mass $n$ and an `electric' (i.e. conventional) mass $m$ can be independently chosen. The more general solution is conveniently written in the following form \cite{Page:1979aj}
\be\label{eq:taubbolt}
ds^2 = V^{-1} \left(d \tau + \sigma \, n  \cos \theta d\phi \right)^2 + V dr^2 + (r^2 - \qtr n^2) \left(d\theta^2 + \sin^2 \theta \, d\phi^2 \right)\,,
\ee
with
\be
V = \frac{r^2 - \qtr n^2}{r^2 - m r + \qtr n^2}\,.
\ee
The Taub-NUT instanton (\ref{eq:tn2}) is recovered by putting $n = m$ and performing a simple change of coordinates. This explains the clumping instability we found above: for centers with the same sign NUT charge, the magnetic and electric gravitational forces cancel, but for centers with different sign NUT charge, both forces are attractive and they add. Thus on average there is a net attraction between the centers.

One can imagine that if there exist instantons with a larger magnetic than electric mass, then these might be able to proliferate and screen gravity. In particular, if instantons existed with magnetic mass but vanishing electric mass then they would behave just like the magnetic monopoles of Polyakov's computation. This is found to be impossible within Einstein gravity, as we now review. The metric (\ref{eq:taubbolt}) has a curvature singularity at $r = \half n$. In order for the solution to be regular, the range of the $r$ coordinate must terminate before reaching this point. This is achieved if the pole of $V$ at larger radius, $r_+$, occurs at $\half n < r_+$. The resulting range of the radial coordinate is $r_+ < r <\infty$, excluding the singular point. This condition on the pole is seen to require $n < m$. That is, regularity of the instanton precisely requires that the electric mass be bigger than the magnetic mass.

In fact, once $\half n < r_+$ one must also require the absence of a conical singularity at the `bolt' at $r=r_+$. This is found to impose $m = \frac{5}{4} n$. The resulting regular solution is called the Taub-bolt instanton \cite{Page:1979aj}. These instantons have a larger mass at fixed NUT charge than the Taub-NUT instantons, and therefore will have even less of a tendency to proliferate.

The upshot of our discussion so far is that interactions between gravitational instantons only amplify the pre-existing nonperturbative instability of hot flat space \cite{Gross:1982cv} and show no counter-acting tendency to proliferate. Hot flat space is furthermore known to be unstable already at a perturbative level. This is essentially the Jeans instability: a finite temperature leads to a thermal population of gravitons that then clump under their gravitational attraction. Mathematically this appears in the fact that gravity anti-screens, so that the graviton acquires a tachyonic thermal mass at one loop \cite{Gross:1982cv}. This perturbative instability may at first sight appear to negate the usefulness of searching for novel nonperturbative physics. However, if a process opposing the seemingly inexorable gravitational clumping can be identified, there may be regimes in which this process could compete with or dominate the thermal instabilities we have just described. We return to this possibility in the discussion section.

The zero temperature compactification of gravity on a spatial circle is also well known to suffer from an analogous nonperturbative instability: the nucleation of `bubbles of nothing' \cite{Witten:1981gj}. States with arbitrarily negative energy exist, e.g. \cite{Brill:1989di, Brill:1991qe}. Also similarly to hot flat space, compactification on a circle leads to perturbative instabilities. In this case the one loop effective potential for the size of the circle causes the circle to either shrink down to the Planck scale or to decompactify, e.g. \cite{Appelquist:1982zs, Rubin:1983ap}. With additional matter content in the theory, such as fermions, the effective potential for the size of the compactified circle can be stabilized, e.g. \cite{Rubin:1983zz}. At a stabilized point, the radion field $\phi$ in the action (\ref{eq:phif}) will acquire a mass term. It was suggested in \cite{Gross:1983mq} that, consequently, at long distances this field would not lead to an attractive force between all instantons, and Polyakov's mechanism of monopole proliferation could occur. In the following section \ref{sec:theorem} we will argue that the correct way to address this question is to ask whether gravitational instantons with NUT charge obey a certain positive mass theorem. If the matter sourcing the spacetime obeys a natural energy condition, we will show that the attractive force between instantons with NUT charge is always larger than the repulsive forces between oppositely charge centers. If this is the case then proliferation cannot occur.

\section{Mass and NUT charge of gravitational instantons}
\label{sec:theorem}

We saw in the previous section that the interactions between well separated gravitational instantons depend on the mass and NUT charge (magnetic mass) of the single center solutions. The reason that the instantons do not proliferate in Einstein gravity is that the attractive force due to the mass overwhelms the screening possibilities of the attractive and repulsive forces due to the NUT charge. Proliferation, at least over some intermediate regime of distance scales, would require that the NUT charge of the single centers be greater or equal to their mass. In this section we ask, beyond pure Einstein gravity, whether for a regular gravitational instanton with NUT charge $n$ and mass $m$ -- with $m,n$ defined carefully below, but in particular agreeing with the vacuum solution (\ref{eq:taubbolt}) asymptotically -- this cannot happen. Namely, regular instantons necessarily satisfy
\be\label{eq:magbound}
m \geq |n| \,.
\ee
We will show that such a bound indeed holds, so long as the matter sourcing the curvature of the spacetime obeys a Euclidean local energy condition.

Positive mass theorems are fundamental results in classical general relativity that underpin, for instance, the stability of Minkowski space \cite{Schon:1979rg}. We are going to prove a slightly different type of positive mass theorem. In particular, we are after a statement about Euclidean gravitational instantons (but we are {\emph{not}} after a positive action theorem). In section \ref{sec:spinors} below we will prove 
(\ref{eq:magbound}) by adapting the spinorial methods of Witten \cite{Witten:1981mf}, and their generalization to include horizons \cite{Gibbons:1982jg}, to Euclidean signature.

Encouraged by this result, in section \ref{sec:PG} we have also attempted, assuming spherical symmetry, to find a Euclidean Penrose-Gibbons inequality \cite{Penrose:1973um, Gibbons:1982wv} with NUT charge and thereby obtain a second derivation of the NUT-Bogomolny bound (the positive energy theorem) without spinors. We have not been successful here. However, having set up the spherically symmetric equations of motion, we have numerically found nontrivial corroboration of our general result.
  
\subsection{Euclidean local energy conditions}

The instantons satisfy the Einstein equations
\be\label{eq:ema}
G_{\mu \nu} = \k^2 T_{\mu \nu} \,.
\ee
We can recall that the usual {\it Lorentzian} positive mass theorems impose the local requirement on the energy-momentum tensor that
\be\label{eq:localbound}
n_\mu n_\nu T_L^{\mu\nu} \geq 0 \,.
\ee
Here $n$ is the unit normal to a constant time slice. This statement is a corrollary of the dominant energy condition.

The Euclidean version of (\ref{eq:localbound}) that arises naturally in the spinorial approach is that, given the normal $n$ to the `time' slice, then
\be\label{eq:ebound}
n_a n_b T^{ab} \leq 0 \,.
\ee
Unfortunately, analytic continuation does not imply that any Lorentzian theory satisfying (\ref{eq:localbound}) has a Euclidean energy-momentum tensor satisfying (\ref{eq:ebound}). This is because each component of the energy-momentum will typically involve contractions of indices, and the contractions of timelike and spacelike indices behave differently under analytic continuation. In particular, considering the cases of a scalar and a Maxwell field, it is easily seen that (i) time dependence and (ii) real electric fields work against the Euclidean bound (\ref{eq:ebound}), while working towards the Lorentzian bound (\ref{eq:localbound}). Thus in requiring (\ref{eq:ebound}) we will be effectively imposing that the matter fields supporting the instantons are not time dependent and do not involve real electric fields. For spatial gradients, magnetic fields and potentials for scalar fields, the Lorentzian and Euclidean bounds have the same content. These complications are not related to NUT charge but are rather present already in trying to prove a positive mass theorem for gravitational instantons. Indeed, it is easily seen that regular Euclidean Reissner-Nordstr\"om black holes with real electric fields can have arbitrarily negative mass at a given fixed charge. This brings us to an issue in semiclassical quantum gravity that we now discuss.

Studies of Euclidean Reissner-Nordstr\"om black holes typically consider pure imaginary electric fields. These are the solutions obtained by simple Wick rotation of the Lorentzian solutions. This class of solutions satisfy our local positive mass bound (\ref{eq:ebound}) and therefore, in agreement with our theorem, the black holes have positive mass. Furthermore, crucially, such black holes can also have positive specific heat. The Euclidean solutions with real electric fields, in contrast, have negative specific heat. While the rules of the semiclassical quantum gravity game are not totally clear, this thermodynamic instability suggests that the solutions will have negative Euclidean modes and that therefore their role in the path integral will be to mediate tunneling (like the neutral Euclidean Schwarzschild black hole \cite{Gross:1982cv}) rather than to proliferate in an instanton gas. It is conceivable that all Euclidean solutions that violate the local bound (\ref{eq:ebound}) due to real electric fields or time dependence also have a negative specific heat or other thermodynamic instabilities. It would be interesting to establish, disprove or otherwise elucidate this statement. For our current purposes of studying the dynamics of NUT charge, we note that the example of negative mass Euclidean Reissner-Nordstr\"om suggests that violation of the local bound (\ref{eq:ebound}) may be associated with thermodynamic instabilities rather than proliferation.

The usual positive mass theorems are formulated as constraints on initial data. Because they bound conserved charges, they therefore hold for all times. Because we are after a Euclidean result, it is less natural to think in terms of initial data. However, we can still (at least locally) take a constant `time' slice of a compactified direction in the geometry. This slice then picks out the `time' direction used in the statement (\ref{eq:ebound}). Note that we are not after a positive action theorem, but a positive mass theorem. It is well known that positive action and positive mass theorems fail for spacetimes with nontrivial asymptotic topology, e.g. \cite{lebrun, Witten:1981gj}. From the spinorial perspective we shall consider first, this has to do with the choice of spin structures in non-simply connected spacetimes \cite{Witten:1981gj}. However, in the positive mass theorem we are after, the constant time slices we use do not contain the compactified direction but rather intersect it and are simply connected. We will see however that counterexamples to the theorem exist if the asymptotically locally flat boundary conditions on the slice are relaxed. Our boundary conditions are fixed by the requirement that we are considering a dilute gas of instantons in the vacuum.

Before moving on to derive the bound (\ref{eq:magbound}), we should note that this bound will prevent instanton proliferation, as we have explained, if gravity is the only long range force that is active. If the instantons carry additional local charges, then the corresponding additional forces may potentially overcome gravitational clumping. Stronger Euclidean positive mass theorems would have to be established in these cases to show otherwise, possibly along the lines of \cite{Gibbons:1982jg}. In the following section \ref{sec:EM} we will discuss the case of Einsten-Maxwell theory. A notable fact -- to be reviewed below -- is that the Taub-NUT instanton admits a normalizable harmonic (anti-)self-dual two form. It can therefore be dressed with electromagnetic fields without these backreacting on the metric.

\subsection{Spinorial proof of positive mass theorem}
\label{sec:spinors}

Recall that we are working with Euclidean signature $(++++)$. We will
take a representation of the Clifford algebra, $\{\gamma^a, \gamma^b\}
= 2 \delta^{ab}$, with all $\gamma$s hermitian. We denote higher rank
elements of the Clifford algebra, i.e.~antisymmetrized products of the
$\gamma^a$, as $\gamma^{ab\dotsb}$; for example $\gamma^{ab} =
\gamma^{[a}\gamma^{b]}$. Here Latin indices refer to components in an
orthonormal basis and shortly we will use middle Latin indices,
e.g.~$i,j,k$, etc., to refer to orthonormal indices on a three
dimensional hypersurface. When needed, Greek indices will refer to
coordinate components.

  In this section we will run through a Euclidean signature version of the standard spinorial proof of the positive mass theorem \cite{Witten:1981mf}. We follow the structure of the argument in \cite{Gibbons:1982jg, Parker:1981uy}. In Lorentzian signature, various previous works have incorporated NUT charge into discussions of positive energy theorems \cite{Gibbons:1984hy, Kallosh:1994ba, Hull:1997kt, Argurio:2008zt}. The upshot of these papers is that in the presence of Lorenztian NUT charge $N_L$, the positive energy result $E_\text{ADM} \geq 0$ gets replaced by $E_\text{ADM}^2 + N_L^2 \geq 0$. One might hope that the Euclidean result is the simple analytic continuation $N_L \to i N$, resulting in our desired bound $E_\text{ADM} \geq |N|$. Indeed, we will see that this is the case, although there is one subtle step. While the structure of the proof is very similar to the Lorentzian case, there are a few differences in the Euclidean argument due to the different gamma matrices involved.

  We consider a four dimensional Euclidean spacetime $M$ and imagine
  we can pick an `initial time slice', i.e.~a three dimensional
  hypersurface $\Sigma$ in $M$. We will return to the question of the global existence of such a slice at the end.
  The ADM momentum is then obtained as a
  surface integral over this hypersurface $\Sigma$. Picking a normal vector $n$ to the hypersurface $\Sigma$, we
  can decompose the spinor covariant derivative, denoted by $\nabla$,
  on $(M,g)$ into a piece intrinsic to $(\Sigma,h)$ and a piece
  depending on the extrinsic curvature:
  \begin{align}
    \nabla_{i} \psi &= \DD_{i} \psi + \frac{1}{2}
    K_{ij} \gamma^{j} \gamma^{4} \psi \, .
    \label{eq:decomp}
  \end{align}
  Here $\DD$ is the covariant derivative on $(\Sigma,h)$ and $K$ is
  the extrinsic curvature. This decomposition is obtained by taking
  the unit normal vector $n$ to be an element of an orthonormal basis,
  so we can take $n_a = \delta_a^4$ in orthonormal indices. This
  implies the projection operator $h_{ab}$ simply projects onto
  indices $i,j=1,2,3$.
  
 One now imposes that the
  spacetime admit a spinor that satisfies a first order linear
  differential equation (the Witten equation) while approaching an
  arbitrary, constant spinor $\psi_0$ at infinity. The Witten operator
  is defined as the Dirac operator constructed from the full space
  covariant derivative on the hypersurface, i.e.~$h_{ab}
  \gamma^a\nabla^b$. The corresponding Witten equation is then $h_{ab}
  \gamma^a \nabla^b \psi = \gamma^i \nabla_i \psi = 0$. Note that
  using the decomposition \Eqref{decomp}, we can write:
  \begin{align}
    \gamma^i \DD_i \psi &= \gamma^i \nabla_i \psi - \frac{1}{2} K
    \gamma^4 \psi\, .
  \end{align}
  These relations tell us that the standard Dirac operator on $\Sigma$
  satisfies:
  \begin{align}
    \left( \gamma^i \DD_i \right)^2 \psi &= \left( \gamma^i \nabla_i
      \psi \right)^2 \psi + \frac{1}{2} \gamma^4 \gamma^i \left( \DD_i
      K \right) \psi - \frac{1}{4} K^2 \psi\, .
  \end{align}
  In addition, the Dirac operator satisfies the usual Lichnerowicz
  identity $(\gamma^i \DD_i)^2 = \DD^i \DD_i - \frac{1}{4}
  R^{(\Sigma)}$, where $R^{(\Sigma)}$ is the Ricci scalar of
  $(\Sigma,h)$. Combining these two results gives:
  \begin{align}
    \DD^i \DD_i \psi &= \frac{1}{4} \left( R^{(\Sigma)} -
      K^2 \right) \psi + \left( \gamma^i \nabla_i \right)^2 \psi +
    \frac{1}{2} \gamma^4 \gamma^i \left( \DD_i K \right)\psi\, .
  \end{align}
  
  If we now consider the divergence of the current $\psi^\dagger
  \nabla^i \psi$, then the above relations give:
  \begin{align}
    \DD_i \left( \psi^\dagger \nabla^i \psi \right) 
    &= \left|\nabla \psi \right|^2 + \psi^\dagger \DD_i \DD^i \psi -
    \frac{1}{4} K_{ij} K^{ij} \psi^\dagger \psi + \frac{1}{2} \left(
      \DD_i K^i_{\hph{i} j} \right) \psi^\dagger \gamma^j \gamma^4
    \psi + K_{ij} \psi^\dagger \gamma^i \gamma^4 \DD^j \psi \nonumber \\
    &= \left|\nabla \psi \right|^2 + \frac{1}{4} \psi^\dagger
    \left[R^{(\Sigma)} - K^2 + K_{ij} K^{ij} + 2 \DD_i \left(
        K^i_{\hph{i} j} + \delta^i_j K\right)
      \gamma^j \gamma^4  \right]\psi \nonumber \\
    &\qquad + \psi^\dagger \left( \gamma^i \nabla_i \right)^2 \psi +
    K_{ij} \psi^\dagger \gamma^i \gamma^4 \nabla^j \psi\, .
    \label{eq:jdiv}
  \end{align}
  The (Euclidean) Gauss-Codazzi relations and Einstein's equations (\ref{eq:ema}) simplify the
  expressions in the middle line of the previous equation, allowing them to be rewritten in
  terms of the stress tensor:
  \begin{align}
    \kappa^2 n_a n_b T^{ab} &= -\frac{1}{2}
    \left[R^{(\Sigma)} - K^2 + K_{ij} K^{ij} \right] \,, &
    \kappa^2 h^j{}_{b} n_c T^{bc} &= \DD_a \left(K^{aj} -
      \delta^{aj} K \right) \, .
  \end{align}
  Using these simplifications, the goal is to argue that \Eqref{jdiv}
  is positive definite with the assumption of a suitable energy
  condition. The first step in this procedure is to pick $\psi$ to
  satisfy $\gamma^i \nabla_i \psi = 0$, i.e.~be a solution of the
  Witten equation. However, in this Euclidean setting, there are still
  some issues we need to worry about. 

  First, and most notably, there is the last term in \Eqref{jdiv},
  which arises due to the hermiticity of $\gamma^4$ (rather than the
  anti-hermiticity of $\gamma^0$ in the Lorentzian proof). This term
  is not manifestly positive so we need to get rid of it. We now show
  that this term is a linear combination of the Witten equation and its
  derivative. Note that:
  \begin{align}
    \DD_i \left( \psi^\dagger \gamma^i \gamma^j \nabla_j \psi \right)
    &= \left(\DD_i \psi\right)^\dagger \gamma^i \gamma^j \nabla_j \psi
    + \psi^\dagger \gamma^i \gamma^j \DD_i \nabla_j \psi \\ 
    &= \left(\DD_i \psi\right)^\dagger \gamma^i \gamma^j \nabla_j \psi
    + \psi^\dagger \left(\gamma^i \nabla_i \right)^2 \psi - K_{ij}
    \psi^\dagger \gamma^i \gamma^4 \nabla^j \psi - \frac{1}{2} K
    \psi^\dagger \gamma^4 \gamma^j \nabla_j \psi\, . \nonumber
  \end{align}
  Assuming $\psi$ solves the Witten equation, we are left with
  $\DD_i \left( \psi^\dagger \gamma^i \gamma^j \nabla_j \psi \right) =
  - K_{ij} \gamma^i \gamma^4 \nabla^j \psi$. Thus, we can add this to
  our expression in \Eqref{jdiv} to cancel the unwanted term:
  \begin{align}
    \DD_i \left[ \psi^\dagger \left( \nabla^i + \gamma^i \gamma^j
        \nabla_j \right) \psi \right] &= \left|\nabla \psi \right|^2 +
    \frac{1}{2} \kappa^2 \psi^\dagger \left[-n_a n_b T^{ab} + h_{ja}
      n_b T^{ab} \gamma^j \gamma^4 \right]\psi\, .
    \label{eq:newjdiv}
  \end{align}
  Because $\psi$ satisfies the Witten equation, the new additional term on the left hand side of
  the above expression is in fact zero. We were inspired to add this term by a similar, mathematically
  natural, term that is present in \cite{Parker:1981uy}.
  
  Second, we note that the $\gamma^j \gamma^4$
  term \Eqref{newjdiv} is purely imaginary, and therefore, upon taking
  the real part, drops out. This isn't an issue so much as something that
  will simplify (relative to the Lorentzian case) the constraints we want to impose on $T_{ab}$ in order
  that the mass be positive.

  All in all, assuming that $\psi$ satisfies the Witten equation and that the matter satisfies the Euclidean energy condition (\ref{eq:ebound}), we
  are left with:
  \begin{align}
    \DD_i \left[ \psi^\dagger \left(\nabla^i +\gamma^i \gamma^j
        \nabla_j \right)\psi + \text{c.c.} \right] &= \left|\nabla
      \psi \right|^2 - \kappa^2 n_a n_b T^{ab} \psi^\dagger \psi \geq
    0\, .
    \label{eq:jdivf}
  \end{align}
  
  Moving on, we now look at the boundary term obtained by integrating
  \Eqref{jdivf} over $\Sigma$:
  \begin{align}
    \int_\Sigma{\DD_i \left( \psi^\dagger \nabla^i \psi + \text{c.c}\right)} &=
    \int_{\partial \Sigma}{ v_i\psi^\dagger \nabla^i \psi + \text{c.c.}}\, ,
    \label{eq:bdy}
  \end{align}
  where $v_i$ is the unit normal to $\partial \Sigma$ in $\Sigma$. Asymptotically we will
  have $\psi \sim \psi_0 + {\cal O}(r^{-1})$. Using
  the Witten equation, we can write:
  \begin{align}
    v_i \psi^\dagger \nabla^i \psi &= v_i \psi^\dagger
    \left(\nabla^i - \gamma^i \gamma^j \nabla_j \right)\psi = - v_i
    \psi^\dagger \gamma^{ij} \nabla_j \psi = - v_i \psi^\dagger
    \gamma^{ij} \left(\DD_j + \frac{1}{2} K_{jk} \gamma^k \gamma^4
    \right) \psi\, .
    \label{eq:bdysimp}
  \end{align}
  Using $\gamma^{ij} \gamma^k = \gamma^{ijk} + 2 \gamma^{[i}
  \delta^{j]k}$ and the symmetry of $K_{ij}$, the terms involving the
  extrinsic curvature can be written:
  \begin{align}
    - \frac{1}{2} v_i K_{jk} \psi^\dagger \gamma^{ij} \gamma^k
    \gamma^4 \psi = \frac{1}{2} v_i \left( K^i{}_{j} - K
      \delta^i_j \right) \psi_0^\dagger \gamma^j \gamma^4 \psi_0 +
    {\cal O}\left(r^{-3}\right)\, ,
    \label{eq:admpi}
  \end{align}
  which is proportional to the ADM spatial momentum density. As for
  the term involving $D_j$, we find (ignoring the $\partial_j$ terms
  as they are subleading):
  \begin{align}
    \gamma^{ij} \DD_j \psi &\simeq \frac{1}{4} \omega_{klj} \gamma^{ij}
    \gamma^{kl} \psi_0 = \frac{1}{4} \omega_{klj} \gamma^{ijkl} \psi_0
    - \omega^j{}_{l j} \gamma^{li} \psi_0+ \frac{1}{2}
    \omega^{ji}{}_{j} \psi_0\, .
  \end{align}
  Therefore upon taking the real part of everything, we find the
  integrand on the right hand side of \Eqref{bdy} becomes:
  \begin{align}
    v_i \psi_0^\dagger \nabla^i \psi_0 + \text{c.c.} &= v_i
    \omega^{ij}{}_{j} \psi_0^\dagger \psi_0 - \frac{1}{2}
    \epsilon^{ijkl} v_i \omega_{jkl} \psi_0^\dagger \gamma^5
    \psi_0 
    \, .
    \label{eq:bdyintg}
  \end{align}
  We note that the term with the ADM spatial momentum has vanished upon
  taking the real part. This is consistent with the fact that a corresponding term
  previously vanished in the bulk integrand upon taking the real part. We defined
  $\gamma^5 = \g^1 \g^2 \g^3 \g^4$.
 
  The first term in the integrand \Eqref{bdyintg} is simply the ADM
  energy density. We can see this by picking an orthonormal basis
  asymptotically of the form $\text{e}_{j\nu} = \delta_{j\nu} + \frac{1}{2}
  h_{j\nu}$, where the metric looks like $g_{\mu \nu} = \delta_{\mu
    \nu} + h_{\mu \nu}$. Then using the general expression for the
  spin connection in terms of the orthonormal basis,
  \begin{align}
    \omega^{ab}{}_{\mu} &= \frac{1}{2} \text{e}^{a \nu}
    \left( \partial_\mu \text{e}^b{}_{\nu} - \partial_\nu
      \text{e}^b{}_{\mu} \right) - \frac{1}{2} \text{e}^{b \nu}
    \left( \partial_\mu \text{e}^a{}_{\nu} - \partial_\nu
      \text{e}^a{}_{\mu} \right) - \frac{1}{2} \text{e}^{a \rho} \text{e}^{b
      \sigma} \left( \partial_\rho \text{e}_{c \sigma} - \partial_\sigma
      \text{e}_{c \rho} \right) \text{e}^c{}_{\mu}\, ,
    \label{eq:spincon}
  \end{align}
  we can contract $\omega^{ij}{}_{\mu}$ with
  $\text{e}_j{}^{\mu}$ to find:
  \begin{align}
    \omega^{ij}{}_{j} &= \text{e}^{i\nu} \text{e}_j{}^{\mu}
    \left( \partial_\mu \text{e}^j{}_{\nu} - \partial_\nu
      \text{e}^j{}_{\mu} \right) = \frac{1}{2} \delta^{i\nu}
    \left( \partial_j h_{j \nu} - \partial_\nu h_{jj} \right)\, ,
  \end{align}
  which is exactly proportional to the ADM energy density $E_\text{ADM}$. We emphasize again that this is the ADM energy defined via a constant `time' slice of the four dimensional Euclidean geometry. It is not the ADM energy of the whole instanton viewed as the spatial part of a five dimensional geometry. The Euclidean time circle is not part of the spatial slice. We therefore avoid the complications afflicting positive energy theorems when the asymptotic space is not simply connected.
  
    Finally, we
  turn to the second term in \Eqref{bdyintg}. Using the natural
  orthonormal basis for metrics that are asymptotically of the form in \Eqref{tn2}, i.e.~:
  \begin{align}
   \text{e}_1 &= U^{1/2} \dd r & \text{e}_2 &= U^{1/2} r & \text{e}_3 &= U^{1/2} r
    \sin \theta \dd \phi & \text{e}_4 &= U^{-1/2} \left(\dd \tau + a
    \right)\, ,
  \end{align}
  and $v_i \to \delta_i^1$, a simple calculation of the spin
  connection in this basis gives:
  \begin{align}
    \frac{1}{2} \epsilon^{ijkl} v_i \omega_{jkl} = \frac{1}{2} f_{23}\, .
  \end{align}
Here $f$ is the field strength corresponding to $a$. This quantity is proportional to the NUT charge $N$.

  Putting it all together, we have upon integrating \Eqref{bdyintg}
  over the sphere at infinity,
  \begin{align}
    \int_{\partial \Sigma} v_i \psi_0^\dagger \nabla^i \psi_0 +
    \text{c.c.}  &= 8 \pi \psi_0^\dagger \left(E_{\text{ADM}} - N
      \gamma^5 \right) \psi_0\, .
  \end{align}
  Thus, since the bulk integrand is positive definite, we conclude
  that the eigenvalues of the operator $E_{\text{ADM}} - N \gamma^5$
  must be positive, i.e.~
  \begin{align}
    E_{\text{ADM}} \geq |N|\, .
    \label{eq:EN}
  \end{align}
  This establishes our claim (\ref{eq:magbound}) upon noting that on the asymptotic vacuum solution
  (\ref{eq:taubbolt}) we have
  \be
  N = \frac{n}{2} \,, \qquad E_\text{ADM} = \frac{m}{2} \,.
  \ee
  Because the vacuum solution will pertain far away from the centers, in our dilute instanton gas description, these quantities $n$ and $m$ will indeed determine the forces between instantons, as we discussed above.
   From \Eqref{jdivf}, we see that the bulk integrand vanishes only
  when there are no sources and when $\nabla \psi = 0$, i.e.~the space
  admits a Killing spinor. This is the case of Taub-NUT.

  In the presence of a bolt -- technically defined as a codimension two fixed
  point surface of an isometry \cite{Gibbons:1979xm}, but more generally, as we are not assuming the presence of any symmetries, we mean a two dimensional interior boundary of our
  spatial slice with vanishing mean curvature -- there will be an extra boundary term in
  \Eqref{bdy}. However, going through similar arguments as in
  \cite{Gibbons:1982jg}, again with minor differences due to different reality conditions on the gamma matrices,
  one can impose a projection condition that commutes with the Witten equation to make this boundary contribution
  vanish due to the zero mean curvature of the bolt in $\Sigma$. We demonstrate this fact in the appendix.

The argument above assumes that an appropriate `spatial slice' hypersurface exists with no boundaries except asymptotically and on any interior bolts. Such a hypersurface is easily constructed for the Taub-NUT geometry (\ref{eq:tn2}). For a minimal charge instanton, this spacetime is topologically $\R^4$, with a squashed three sphere at each radius. Each squashed three sphere contains a topological two sphere\footnote{For instance, at fixed radius $r$ in the coordinates of (\ref{eq:tn2}), for e.g.~$n=1$, $\tau = - \phi$ and $\tau = -\phi + 2\pi$ define two topological hemispheres that are joined at $\theta = \pi$ to form a topological two sphere.}. Radially filling in these two spheres, we get a hypersurface with topology $\R^3$. This will do for our theorem. We expect the single centered instantons that we are interested in to have the same topological structure (up to the presence of bolts), and therefore we anticipate that the required hypersurface will exist.

There is a vacuum solution that might appear to fall under the remit of our discussion and which has a negative ADM mass: the Atiyah-Hitchin instanton \cite{Atiyah:1985dv}. Interestingly, this spacetime does not have a $U(1)$ time translation symmetry, consistent perhaps with our previous remarks that time dependence generates negative mass in Euclidean signature.
However, the lack of an isometry is not the main issue. The regularity of the Atiyah-Hitchin instanton requires a certain $\Z_2$ action that antipodally identifies the locally asymptotically flat geometry, see e.g. \cite{Gibbons:1986df}. These boundary conditions are not compatible with our finite temperature instanton gas. Perhaps these instantons play an interesting role in a different context.

\subsection{Penrose-Gibbons inequalities and numerical investigations}
\label{sec:PG}

A different approach to bounding the mass of black holes comes from the Penrose inequality \cite{Penrose:1973um}. This inequality was sharpened for charged black holes by Gibbons \cite{Gibbons:1982wv}. In particular, the Bogomolny bound on the mass by the charge follows from the Penrose-Gibbons inequality. In this section we briefly explore the possibility of re-obtaining our
analogous Bogomolny bound (\ref{eq:magbound}) using similar arguments.

In general it is not simple to prove Penrose-Gibbons inequalities. However, the situation simplifies greatly if spherical symmetry is assumed. In this case the Penrose-Gibbons inequality admits a simple proof \cite{Hayward:1998jj}. Furthermore, it is easily checked that  the arguments in \cite{Hayward:1998jj} apply equally to time-independent, spherically symmetric Euclidean spacetimes. Here will shall adapt the steps in \cite{Hayward:1998jj} to the Euclidean case with NUT charge. We will find that NUT charge complicates the argument and the bound on the mass we are after does not drop out of this approach. We hope that our discussion here will stimulate future work.

Considering spherical symmetry and time independence, however, reduces the problem to solving a small number of ODEs. This allows us to numerically compute the instanton mass for a given NUT charge in the presence of matter sources and to check explicitly if $m \geq n$ holds. We find that indeed $m \geq n$ is satisfied in our numerical examples in a nontrivial way, corroborating our general result.

The appropriate spherically symmetric ansatz for the metric including NUT charge is
\be\label{eq:spherical}
ds^2 = f(r) \left(d\tau + n \cos \theta d \phi \right)^2 + g(r) dr^2 + r^2 \left(d\theta^2 + \sin^2\theta d\phi^2 \right) \,.
\ee
This spacetime has a $U(1) \times SU(2)$ isometry. As previously, the time circle identification and NUT charge are
\be\label{eq:ttemp}
\t \sim \t + \frac{1}{T} \,, \qquad n = \frac{1}{4 \pi T} \,.
\ee
The Asymptotically Locally Flat condition requires $f \to 1$ as $r \to \infty$. We wish to source this spacetime with an energy momentum tensor. A simple energy momentum tensor compatible with the symmetries of the metric is parametrized by two additional functions $\rho(r)$ and $p(r)$. Writing the energy momentum tensor as a symmetric two tensor:
\be\label{eq:tmunu}
T = \rho(r) f(r) \left(d\tau + n \cos \theta d \phi \right)^2 + p(r) \left[ g(r) dr^2 + r^2 \left(d\theta^2 + \sin^2\theta d\phi^2 \right) \right] \,.
\ee
We will not impose any equations of motion on $\rho$ and $p$ beyond the conservation law $\nabla_a T^{ab} =0$. The Einstein (\ref{eq:ema}) and conservation equations on this ansatz reduce to
\bea
\frac{r f'}{f} & = & g -1 - \frac{n^2}{4} \frac{f g}{r^2} + \k^2 r^2 g \, p \,, \\
\frac{r g'}{g} & = & \frac{1}{g} - 1 + \frac{3 n^2}{4} \frac{f}{r^2} - \k^2 r^2 \rho \,, \label{eq:geq}\\
r p' & = & (\rho - p) \frac{r f'}{2f} + 2(g-1) p \,.
\eea
In these equations the energy density $\rho$ is to be chosen freely, so long as it is compatible with the boundary conditions we will want to impose on $f$ and $g$.

The local Euclidean positive energy condition (\ref{eq:ebound}) on the metric and energy momentum tensor above is, taking $n$ to be the unit normal to a locally constant time slice,
\be\label{eq:truebound}
n_a n_b T^{ab} = \frac{r^2 \rho + p \, g \, n^2 f \cot^2 \theta}{r^2 + n^2 f \cot^2 \theta} \leq 0 \,.
\ee
Because this is a little cumbersome to work with, we will instead impose the necessary but weaker condition that $\rho \leq 0$. This is a weaker condition because it does not imply (\ref{eq:truebound}). The weaker condition is sufficient, however,
to corroborate our NUT-Bogomolny bound. We shall see that we do not find any counterexamples to (\ref{eq:magbound}), even with this weaker local energy condition.

For concreteness let us focus on the case in which the Euclidean time circle collapses on a NUT rather than a bolt in the interior. Experience with other cases (such as the Einstein theory results of section \ref{sec:grav}) suggests that inclusion of a bolt is likely to increase the mass at fixed NUT charge. Collapsing on a NUT means that the $r$ coordinate in the metric (\ref{eq:spherical}) goes all the way down to $r=0$ and that, in these coordinates, $g \to 4$ while $f \sim r^2/n^2$ as $r \to 0$ for regularity.
Following \cite{Hayward:1998jj} we can write
\be\label{eq:mexp}
m = \lim_{r \to \infty} r \left(1 - \frac{1}{g} \right) = \int_0^\infty \frac{d}{dr} \left(r - \frac{r}{g} \right) dr = \int_0^\infty \left(\frac{3 n^2}{4} \frac{f}{r^2} - \k^2 r^2 \rho \right) dr \,.
\ee
The first expression for the mass follows, for instance, from using the asymptotic form of the metric (\ref{eq:taubbolt}), which will pertain so long as the matter sources fall off sufficient quickly asymptotically. To obtain the final expression we used the equation of motion (\ref{eq:geq}) in the last step.

In the charged black hole case \cite{Hayward:1998jj}, the analogous expression to (\ref{eq:mexp}) with charge $q$ and horizon radius $r_h$ turns out to be $m = r_h + \int_{r_h}^\infty \left[q^2/(4 r^2) - \k^2 r^2 \rho \right] dr$. Assuming $\rho \leq 0$, this leads to $m \geq r_h + q^2/(4 r_h) \geq |q|$, directly recovering the Bogomolny bound in that case. In the present case, all we can conclude from (\ref{eq:mexp})  is that
\be\label{eq:34}
m \geq \frac{3 n^2}{4} \int_0^\infty  \frac{f dr}{r^2} \,.
\ee
One can indeed check that for Taub-NUT itself -- which in these coordinates corresponds to $f = (n^2 + 2 r^2 - n \sqrt{n^2 + 4 r^2})/(2r^2)$ -- the integral above gives precisely $n$, as it should. However, it is unclear how to turn this fact into a general bound that is saturated by Taub-NUT. In fact, in our numerical examples that follow, we will find cases for which, while satisfying $m > n$, the integral on the right hand side of (\ref{eq:34}) is not greater than $n$, so that the full integral expression in (\ref{eq:mexp}), including the energy density term, is needed.

To search numerically for violation of $m \geq n$ we solve the equations of motion above for $\{f,g,p\}$, given some function $\rho$ as an input. We find the following series expansion for regular solutions at the NUT
\bea
f & = & \frac{r^2}{n^2} + \frac{g_2 \, r^4}{4\, n^2} + \cdots \,, \\
g & = & 4 + g_2 r^2 + \frac{(7 g_2^2 - 96 \k^2 \rho_2) r^4}{16} + \cdots \,, \label{eq:gexp} \\
p & = & -\frac{\rho_2 r^2}{3} + \frac{(g_2 \rho_2 - 3 \rho_4) r^4}{3} + \cdots \,.
\eea
There is a single constant of integration $g_2$ in addition to the general form for the energy density $\rho = \rho_2 r^2 + \rho_4 r^4 + \cdots$. For accurate numerics we use this expansion to high order. We will choose two forms for the energy density
\be\label{eq:rhoform}
\rho(r) = - \a \, r^2 e^{- (r/r_o)^2} \,, \qquad \rho(r) = - \frac{\a \, r^2}{1 + (r/r_o)^6} \,.
\ee
Both of these functions are characterized by two constants $\{\a,r_o\}$. We take $\a > 0$ so that the matter obeys the (weakened) local Euclidean energy condition $\rho < 0$ discussed above. On the solutions we find, $p$ will be positive in places and so the stronger local condition (\ref{eq:truebound}) will not hold everywhere. We will find that $m \geq n$ despite this fact.

The solutions can be found by numerical shooting. Fix the NUT charge $n$ from the outset (this corresponds to fixing the temperature via (\ref{eq:ttemp})). One finds that for a given $\{\a,r_o\}$ there are two possibilities. Over some range of these parameters there exists a value of the constant $g_2$ in (\ref{eq:gexp}) such that $f \to \text{const.}$ as $r \to \infty$, while for the remainder of the regime of parameters, no such solution is found. However, for asymptotic local flatness we need to impose the stronger condition that $f \to 1$. At this point we can note that given a solution $\{ f(r),g(r),\r(r),p(r) \}$ to the equations of motion, then $\{\l^2 f(r/\l),g(r/\l), \r(r/\l)/\l^2,p(r/\l)/\l^2\}$ is also a solution for any constant $\l$. Therefore, once we have found a solution such that $f \to \text{const.}$ asymptotically, we can find a solution with the correct boundary conditions by rescaling. This rescaling effectively changes the constants $\{\a,r_o\}$, leaving the combination $\a r_o^4$ invariant. The upshot is that we find a solution with the correction boundary conditions for a range of values of $\a r_o^4$. Given a solution, the mass can then be read off using (\ref{eq:mexp}).

\begin{figure}[h]
\begin{center}
\includegraphics[height=200pt]{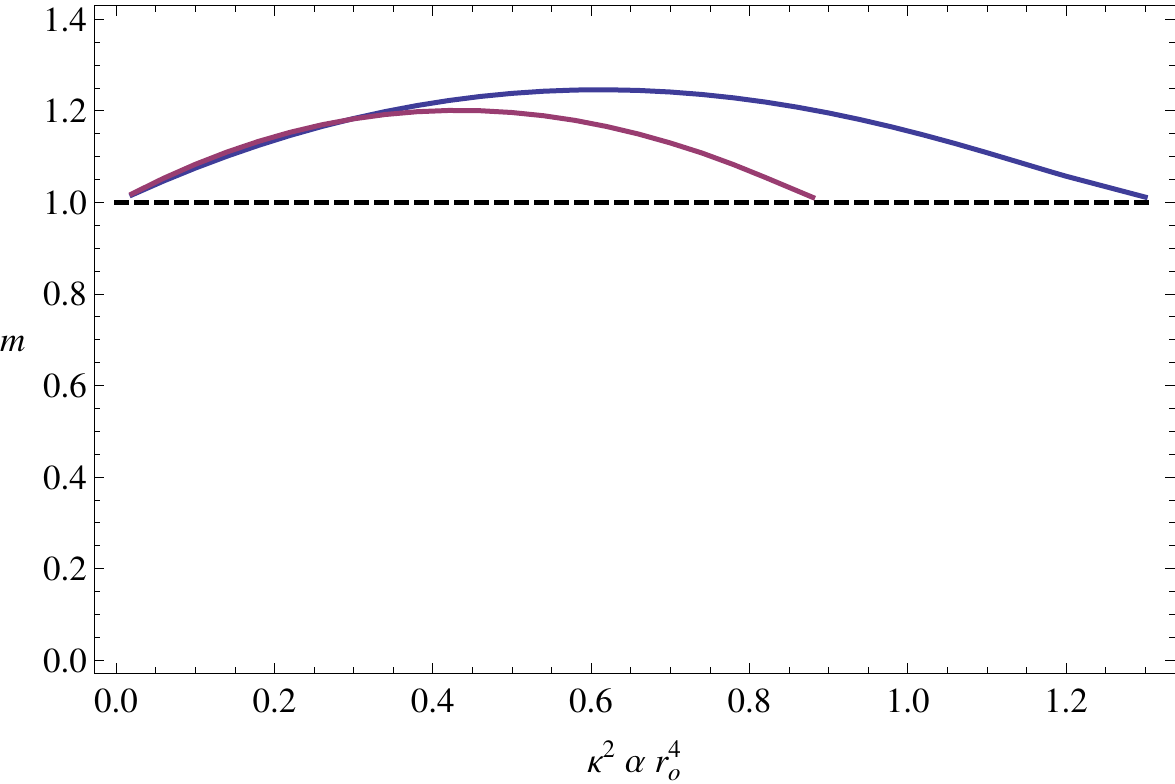}
\caption{The mass of two families of spacetimes with energy densities (\ref{eq:rhoform}) as a function of the parameter $\a r_o^4$. The curves are shown over the entire parameter range for which solutions exist. The top curve corresponds to the exponential form of the density in (\ref{eq:rhoform}). The NUT charge $n=1$ is shown with a dashed line. We see that $m \geq n$ for all solutions found and that each family of solutions exhibits a maximal mass.
\label{fig:masses}}
\end{center}
\end{figure}

In figure \ref{fig:masses} below we plot the mass of the solutions found as a function of $\a r_o^4$ for each of the two functions chosen in (\ref{eq:rhoform}).
The plot is shown over the whole range of $\a r_o^4$ for which the solutions exist. The plot shows two interesting features. Firstly, there is a maximal mass. This is a manifestation of the familiar bound on the mass of stars in general relativity (although we have specified the energy distribution rather than the more usual equation of state). Solutions above this mass presumably necessarily contain a bolt rather than a NUT. Secondly, in beautiful agreement with our general result, $m \geq n$ over the whole range where the solutions exist. The bound $m = n$ is saturated in the limit as $\a \to 0$ and the matter becomes negligible (in this limit the solution becomes Taub-NUT) and also at the maximal value of $\a$, beyond which no solutions exist. This latter fact in particular seems to be a nontrivial corroboration of the bound $m \geq n$. As we noted below equation (\ref{eq:34}) above, the system achieves this bound in a delicate way. We find by explicitly checking on the solutions that one cannot drop the energy density term in rightmost expression in (\ref{eq:mexp}) as one would in establishing the Bogomolny bound for charged black holes.

\section{Einstein-Maxwell theory}
\label{sec:EM}

The considerations of the previous section are not the whole story if the NUT charged instantons can be dressed with additional fields. This will result in additional forces between instantons, beyond those due to the electric and magnetic gravitational masses. An interesting explicit case is to add a Maxwell field to pure gravity. This theory has the interesting feature that the NUT centers can be dressed with certain electromagnetic fields without any explicit sources. Our original motivation for considering Einstein-Maxwell theory was the existence of exact multi-instanton solutions that contain centers with both NUT and anti-NUT charges \cite{Yuille:1987vw, Whitt:1984wk, Dunajski:2006vs}. This is distinct from pure gravity -- in which the only exact multi-instanton solutions are the Gibbons-Hawking metrics with either all NUT or all anti-NUT centers \cite{Gibbons:1979zt} -- and suggested the possibility of NUT charge proliferation in Einstein-Maxwell theory. The results in this section, however, will indicate that in the path integral these exact solutions are overwhelmed by gravitational and electromagnetically induced clumping. The exact solutions may possibly play a preferred role in a supersymmetric context \cite{Dunajski:2006vs}.

Including a Maxwell field, the Euclidean action now becomes
\be\label{eq:em}
S_E = \int d^4 \sqrt{g} \left(-\frac{1}{2 \k^2} R + \frac{1}{4 e_4^2} F^2 \right) - \frac{1}{\k^2} \int d^3x \sqrt{\gamma} K \,.
\ee
The corresponding equations of motion are
\be
R_{\mu\nu} - \half R g_{\mu\nu} = \frac{\k^2}{e_4^2} \left(F_{\mu\s}F_\nu{}^\s - \qtr g_{\mu\nu} F^2 \right) \,, \qquad \nabla_\mu F^{\mu\nu} = 0 \,.
\ee
The single-centered instantons will again have a metric given by the Euclidean Taub-NUT geometry (\ref{eq:tn2}).
Famously \cite{Eguchi:1978gw, Pope:1978zx, Gauntlett:1996cw, Lee:1996if, Gibbons:1996wc}, the Taub-NUT solution admits a normalizable harmonic two form, with no explicit sources. Specifically, the metric may be dressed with the gauge potential
\be\label{eq:asingle}
A = \frac{k}{U} \left(d \tau + \sigma \, n  \cos \theta d\phi \right) \,.
\ee
Because the harmonic field strength $F = dA$ is self or anti-self dual,
\be
\star F = \sigma F \,,
\ee
the Maxwell energy-momentum tensor vanishes and the metric (\ref{eq:tn2}) is unchanged. In particular, the real electric fields in this case are compatible with the local bound (\ref{eq:ebound}) and do not lead to negative mass or thermodynamic instabilities. The solution now carries both electric and magnetic charge of the Maxwell field. Somewhat exotically, there is no source for the electromagnetic field and furthermore no homologically nontrivial two cycles in the geometry. Therefore through any closed two cycle $ \int_{\Sigma} F = 0$, and the magnetic flux is not quantized \cite{Pope:1978zx}. The underlying NUT charge in the geometry remains quantized and the continuous parameter $k$ in the Maxwell field will not lead to IR divergences in the instanton gas.

The single instanton action is now easily computed to be
\be
s_{o,k} = \frac{1}{2 (e_4 T)^2} \, k^2 + s_o\,.
\ee
The Maxwell action contributes the first of these terms, with the gravitational contribution carrying over from (\ref{eq:tnaction}).

To describe the interaction between multiple instantons, we again use the metric ansatz (\ref{eq:metricreduce}), together with the Maxwell field
\be
A = \psi \, d\tau + b \,.
\ee
Here the electromagnetic potentials $\psi$ and $b$ are independent of $\tau$. For well-separated instantons, the interactions again occur in a regime where the fields $\{\phi,\psi,a,b\}$ may be linearized. The action (\ref{eq:em}) evaluated on this ansatz and to quadratic order now becomes
\be
 S_E^{(2)} = \frac{1}{2 \k^2 \, T} \int d^3x \left(\frac{1}{4} f^2 + \frac{1}{2} \left( \nabla \phi\right)^2 \right) +  \frac{1}{e_4^2 \, T} \int d^3x \left(\frac{1}{4} g^2 + \frac{1}{2} \left( \nabla \psi\right)^2 \right) \,.
\ee
The field strength $g = db$. S-dualizing the field strengths and including the instanton sources gives the action
\bea
\lefteqn{\int d^3x \Bigg[\frac{1}{2 \a_o^2} \left( \left(\pa \Phi \right)^2 + \left(\pa \Psi \right)^2 \right) + \frac{1}{2 \b_o^2} \left( \left(\pa \rchi \right)^2 + \left(\pa \Theta \right)^2 \right)} \nonumber\\
 && - \delta^{(3)}(x) \Big(\left[i \sigma \Phi(x) + \Psi(x) \right] + k \, \left[i \sigma \rchi(x) + \Theta(x) \right]\Big)  \Bigg] \,.
\eea
Here $\rchi$ is the dual to $g$ and we rescaled $\psi$ to obtain $\Theta$. We introduced
\be
\b_o^2 = \frac{1}{e_4^2 \, T^3} \,.
\ee

Once again summing over the multi-instanton contributions we obtain the effective action. In this case we must also integrate over the electromagnetic charge $k$ of each instanton as well as the positions of the centers, as this charge is not quantitized. The integral that arises in performing the sum over instantons is
\be
\int_{-\infty}^\infty dk e^{-k^2/2(e_4 T)^2} e^{k \Theta} \cos \left(\Phi + k \rchi \right) = \sqrt{2 \pi} (e_4 T) e^{(\Theta^2 - \rchi^2) (e_4 T)^2/2} \cos \left(\Phi + (e_4 T)^2 \rchi \Theta \right) \,. \label{eq:integral2}
\ee
Here again we have not been careful with the measure for the $k$ modulus. In particular, the above expression will be multiplied by an undetermined constant. Denoting this coupling by $\zeta$, we have obtained
\bea
\widetilde S_E[\Phi,\Psi,\rchi,\Theta]  & = & \int d^3x \Bigg[ \frac{1}{2 \a_o^2} \left( \left(\pa \Phi \right)^2 + \left(\pa \Psi \right)^2 \right) + \frac{1}{2 \g_o^2} \left( \left(\pa \rchi \right)^2 + \left(\pa \Theta \right)^2 \right) \nonumber \\
 & & - 2 \, \zeta \, e^{-s_o} e^{\Psi} e^{(\Theta^2 - \rchi^2)/2} \cos \left(\Phi + \rchi \Theta  \right) \Bigg] \,. \label{eq:effcharged}
\eea
We allowed ourselves to rescale $\Theta$ and $\rchi$ to remove the factors of $e_4 T$ in (\ref{eq:integral2}) and introduced
\be
\g_o^2 = \frac{1}{T} = 2 \k^2 T^2 \, \a_o^2 \,.
\ee
We see that, after performing the integral over $k$, all the factors of $e_4$ have dropped out of the effective action (up to a possible non-exponential dependence absorbed into $\zeta$).

The potential in the effective action (\ref{eq:effcharged}) remains unbounded below due to the overall factor of $-e^{\Psi}$. This is not surprising: all the instantons experience a gravitational attraction irrespective of their electromagnetic charges. However, even if we were to find a regime in which the gravitational attraction was parametrically weak compared to the electromagnetic interaction, we see that in the electromagnetic sector the term $e^{(\Theta^2 - \rchi^2)/2}$ in the potential is also unbounded below. The upshot is that Einstein-Maxwell theory also does not experience proliferation of NUT charge.

\section{Future directions}

We have proven that, if appropriate local energy conditions hold, semiclassical quantum gravity is self-consistently quantum ordered. By this we mean that the proliferation of NUT charge that could partially screen long range gravitational interactions is suppressed by the universal gravitational attraction. The price that is paid for this robustness of the massless graviton is the ubiquity of perturbative and non-perturbative clumping instabilities in gravity.

Beyond the semiclassical gravitational instantons we have discussed, NUT charge could also be carried by microscopic gravitational magnetic monopoles. See e.g. \cite{Bunster:2006rt} for a recent discussion of gravitomagnetic sources. If these are sufficiently light, then they might be able to proliferate. While, as we have seen in this paper, gravitational physics tends to require that masses be larger than charges  -- the most familiar example being Reissner-Nordstr\"om black holes -- this does not apply to microscopic excitations. For example, the electron has a charge much larger than its mass. It is of interest to understand what physics controls the mass of microscopic gravitational magnetic monopoles, if they can exist.

If NUT charge were to proliferate in a finite temperature context it would, following the computation in section \ref{sec:polyakov}, gap out fluctuations in the gravitomagnetic field $h^a \equiv \epsilon^{abc }\pa_b \delta g_{\tau c}$. These graviton modes couple to momentum sources. The physical consequence would therefore be the screening of velocity-dependent, i.e. frame-dragging, interactions in the theory. The mechanism would not screen static gravitational interactions, at least at a linearized level. It remains to explore nonpertubrative physics that could potentially screen the Newtonian interaction. Indeed, more generally, a comprehensive classification of the possible low energy quantum phases of gravity would be desirable.

We have focussed on generic aspects of instanton physics in quantum gravity. In the case of supersymmetric theories, one might  hope to use localization techniques to perform the sum over gravitational instantons exactly. We noted in our discussion of Einstein-Maxwell theory the existence of supersymmetric asymptotically flat multi-instanton solutions. In a supersymmetric context, perhaps the extra electromagnetic force can be put to work and novel phases of gravity can be achieved. Relatedly, supersymmetric multicenter solitonic Lorentzian solutions with NUT charge can be constructed in five dimensional theories with Chern-Simons interactions -- see e.g. \cite{Gibbons:2013tqa} for a recent discussion. Euclidean cousins of these solutions may potentially play an interesting role in supersymmetric partition functions.

The essential message of this paper has been that gravitational clumping can prevent proliferation of NUT charge, at least in the simplest contexts. It is well known that an expanding universe works against gravitational attraction. For instance, de Sitter space does not suffer the clumping instabilities of hot flat space, despite the thermal nature of the static patch \cite{Ginsparg:1982rs}. It seems possible that in a cosmological context, one might find situations in which gravitational instantons overcome their attractive forces and proliferate (ideally, without diluting themselves into insignificance). From the perspective of a cosmological observer, this could potentially lead to confining gravitational dynamics. Alternatively, within frameworks related to the wavefunction of the universe \cite{Hartle:1983ai, Witten:2001kn, Strominger:2001pn, Maldacena:2002vr, Anninos:2011ui} proliferation of gravitational instantons would presumably lead to a complicated topological structure at future infinity. The topology of future infinity prescribed by the wavefunction of the universe was recently discussed in \cite{Anninos:2012ft, Banerjee:2013mca}. In terms of gravitational computations, these authors considered a simple set of solutions to pure gravity with a cosmological constant. In keeping with the point of view advocated in this paper, we note that non-saddle point multi-instanton geometries may make an important contribution to the wavefunction of the universe. We intend to study this question in the future.

\section*{Acknowledgements}

We are very grateful to Sung-Sik Lee for collaboration and important input in the initial stages of the this project. We have also benefited from helpful exchanges with Dionysios Anninos, Gary Gibbons, Gary Horowitz, Elias Kiritsis, Hong Liu, Don Marolf, John McGreevy, Mukund Rangamani, Jorge Santos and David Tong. SAH is partially supported by a
DOE Early Career Award, a Sloan fellowship and the Templeton foundation. DMR is supported by a Morgridge Family Stanford Graduate Fellowship.

\appendix
\section{Vanishing of the bolt boundary term}
  \label{sec:bolt}
  
  In this appendix we verify that the additional boundary term
  that appears in the spinorial calculation in the presence of a bolt can be made to vanish.
    As discussed in the main text, the boundary terms are of the form
  $\int_{\partial \Sigma} v_i \left[\psi^\dagger \nabla^i \psi +
    \text{c.c.} \right]$. Using the Witten equation, we can write this
  derivative as $v_i \nabla^i \psi = -v_i \gamma^{ij} \nabla_j
  \psi$. To simplify, note that we are free to pick an orthonormal
  basis such that $v_i \to \delta_i^1$ at the bolt and so we decompose
  the covariant derivative $\nabla_i$ in terms of the extrinsic
  curvatures $K$ of $\Sigma$ and $J$ of the bolt in $\partial
  \Sigma$. To do so, we write $\DD_i \psi$ as:
  \begin{align}
    \DD_I \psi &= {\cal D}_I \psi + \frac{1}{2} J_{IJ} \gamma^J
    \gamma^1 \psi\, , 
  \end{align}
  where ${\cal D}$ is the covariant derivative intrinsic to the bolt
  and the capital indices refer to indices tangent to the bolt. Combining this with \Eqref{decomp}, we can write:
  \begin{align}
    v_i \psi^\dagger \nabla^i \psi &= -v_i \psi^\dagger \gamma^{ij}
    \nabla_j \psi \\
    &= \psi^\dagger \gamma^1 \gamma^I {\cal D}_I \psi - \frac{1}{2}
    K_{1I} \psi^\dagger \gamma^I \gamma^4 \psi + \frac{1}{2}
    \psi^\dagger \left[ J + \left(K - K_{11} \right) \gamma^1 \gamma^4
    \right] \psi\, .
  \end{align}
  Taking the real part of this expression leaves:
  \begin{align}
    v_i \left(\psi^\dagger \nabla^i \psi + \text{c.c.} \right) &=
    \psi^\dagger \gamma^1 \gamma^I {\cal D}_I \psi + \left({\cal D}_I
      \psi \right)^\dagger \gamma^I \gamma^1 \psi + J \psi^\dagger
    \psi\, . \label{eq:Jreal}
  \end{align}
  If now impose an exactly analogous projection condition to that in
  \cite{Gibbons:1982jg}, namely:
  \begin{align}
    i \gamma^1 \gamma^4 \psi = \psi\, ,
    \label{eq:projection}
  \end{align}
  then the derivative terms in (\ref{eq:Jreal}) will vanish as $\{\gamma^1 \gamma^4,
  \gamma^1 \gamma^I\} = 0$. It is easily verified that this condition
  is compatible with the Witten equation. 
  
  As in our discussion of the bulk term in the main text, taking the real
  part of the boundary term gave a slightly simpler expression (\ref{eq:Jreal}) than is obtained
  in the Lorentzian version of this argument. After imposing the projection
  (\ref{eq:projection}), the bolt contribution now vanishes because the
  trace of the extrinsic curvature $J$ is zero, because the bolt is a minimal
  surface.

\end{document}